\documentclass[a4paper]{jpconf}
\usepackage{graphicx}
\begin{document}
\title{High-Pressure Electrical Resistivity Measurements of EuFe$_2$As$_2$ Single Crystals}

\author{N Kurita$^{1,2}$, M Kimata$^{1,2}$, K Kodama$^{1,3}$, A Harada$^1$, M Tomita$^1$,\\ H S Suzuki$^1$, T Matsumoto$^1$, K Murata$^4$, S Uji$^{1,2,3}$ and T Terashima$^{1,2}$}


\address{$^1$National Institute for Materials Science, Tsukuba, Ibaraki 305-0003, Japan}
\address{$^2$JST, Transformative Research-Project on Iron Pnictides (TRIP), Chiyoda, Tokyo 102-0075, Japan}
\address{$^3$Graduate\,School\,of\,Pure\,and\,Applied\,Sciences,\,University\,of\,Tsukuba,\,Ibaraki\,305-0003,\,Japan}
\address{$^4$Division of Molecular Materials Science, Graduate School of Science, Osaka City University, Osaka 558-8585, Japan}
\ead{KURITA.Nobuyuki@nims.go.jp}

\begin{abstract}

High-pressure electrical resistivity measurements up to 3.0\,GPa  have been performed on EuFe$_2$As$_2$ single crystals with residual resistivity ratios $RRR$\,=\,7 and 15. At ambient pressure, a magnetic\,/\,structural transition related to FeAs-layers is observed at $T_0$\,=\,190\,K and 194\,K for samples with $RRR$\,=\,7 and 15, respectively. Application of hydrostatic pressure suppresses $T_0$, and then induces similar superconducting behavior in the samples with different $RRR$ values. However, the critical pressure $\sim$\,2.7\,GPa, where $T_\mathrm{0}$\,$\rightarrow$\,0, for the samples with $RRR$\,=\,15 is slightly but distinctly larger than $\sim$2.5\,GPa for the samples with $RRR$\,=\,7. 
\end{abstract}

\section{Introduction}

Since the discovery of superconductivity in LaFeAs(O,F) with $T_\mathrm{c}$\,=\,26\,K\,\cite{Kamihara}, a family of Fe-pnictide superconductors has attracted much attention. In particular, $A$Fe$_2$As$_2$ ($A$\,=\,Ca,\,Sr,\,Ba,\,Eu,\,etc.) with a tetragonal ThCr$_2$Si$_2$-type structure has been intensively studied because of the availability of stoichiometric single crystals with high quality. It turned out that, in Fe-pnictide compounds, the superconducting (SC) ground state could appear in accordance with the suppression of a magnetic/structural transition by doping\,\cite{Ishida_review}. In the phase diagrams, it is argued that the superconductivity could coexist and/or compete with the antiferromagnetism\,\cite{Chen_BaKFe2As2,Wang_NFL}. However, a random potential introduced by doping could smear the intrinsic SC properties. For understanding the origin of the high-$T_\mathrm{c}$ superconductivity with $T_\mathrm{c}$ up to 55\,K\,\cite{ZARen2008a}, it is of considerable importance to probe the systematic change of ground states using high-quality single crystals. An alternative way to tune the ground state is to apply hydrostatic pressure\,($P$). For instance, recent high-$P$ ac-susceptibility and resistivity measurements have revealed that $A$Fe$_2$As$_2$ ($A$\,=\,Sr,\,Eu) exhibits $P$-induced bulk superconductivity by suppressing the magnetic/structural transition\,[6$-$10]. Meanwhile, superconductivity under hydrostatic $P$ is absent in CaFe$_2$As$_2$\,[11$-$14], and remains a controversial issue in BaFe$_2$As$_2$\,[8,\,15$-$17]

Among the $A$Fe$_2$As$_2$ series, EuFe$_2$As$_2$ is quite unique because the localized Eu$^{2+}$ moments order antiferromagnetically at $T_\mathrm{N}\sim$20\,K, in addition to the magnetic/structural transition related to FeAs-layers at $T_0$\,$\sim$\,190\,K\,[18$-$21]. Interestingly, the magnetic order of Eu$^{2+}$ moments can be detected even in the SC state induced by doping or application of pressure, which could be a main reason for the novel reentrant-SC-like behavior\,[9,\,10,\,22$-$25]. 

Here, we report the results of high-$P$ electrical resistivity measurements in EuFe$_2$As$_2$ using newly grown single crystals with a residual resistivity ratio ($RRR$) as high as 15. At ambient $P$, the magnetic/structural transition occurs at a higher temperature of $T_0$\,=\,194\,K, compared with 190\,K for single crystals with $RRR$\,=\,7. Consequently, it is found that the higher quality single crystal requires higher-$P$ to suppress $T_0$, and to induce the SC ground state in EuFe$_2$As$_2$.

\section{Experimental Details}

Single crystals of EuFe$_2$As$_{2}$ were grown by Bridgman method from a stoichiometric mixture of the constituent elements. In this study, we examined several crystals from two different batches with residual resistivity ratios $RRR$\,=\,7 and 15,  where $RRR$ is defined as $\rho_{\mathrm{300K}}/\rho_{\mathrm{4K}}$. Single crystals measured in Ref\,\cite{Terashima_Eu1} were taken from a batch with $RRR$\,=\,7. High-pressure resistivity measurements of samples with $RRR$\,=\,7 and 15 have been performed simultaneously up to 3.0\,GPa using a hybrid-type piston cylinder pressure device\,\cite{PistonCell}. The resistivity was measured by the four-probe method with an ac current $I$\,=\,0.3\,mA in the  $ab$\,-\,plane. To generate hydrostatic pressure, Daphne 7474 (Idemitsu Kosan) oil, which remains in a liquid state up to 3.7\,GPa at room temperature\,\cite{Daphne7474}, was used as a pressure-transmitting medium. Samples were cooled down in Oxford $^4$He system, slowly with an average rate of 0.5\,K/min. Applied pressure was estimated at 4.2\,K from the resistance change of a calibrated Manganin wire\,\cite{Terashima_erratum}. 


\section{Results and Discussions}
\begin{figure}[b]
\begin{center}
\includegraphics[width=0.65\linewidth]{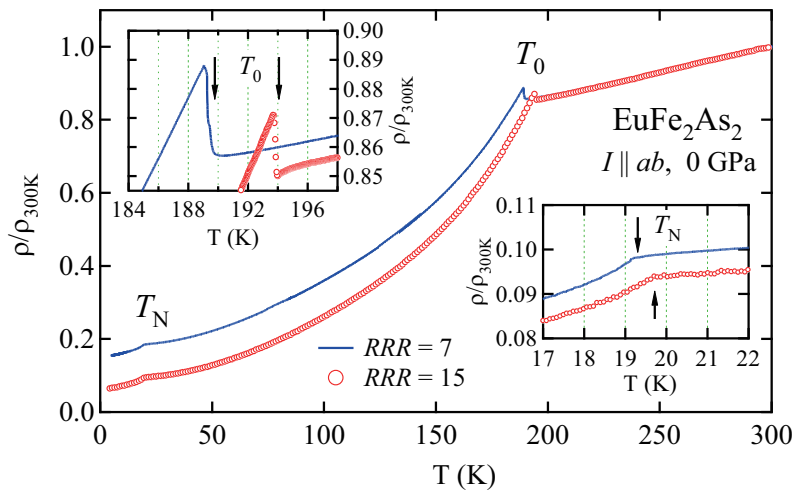}
\end{center}
\caption{(Color online) The scaled electrical resistivity  $\rho/\rho_\mathrm{300K}$ versus temperature in EuFe$_2$As$_{2}$ single crystals  with $RRR$\,=\,7 and 15. The measurement was carried out in zero-field at ambient pressure with the current direction $I$\,$\parallel$\,$ab$. Upper left and lower right insets represent the expanded views around $T$\,=\,$T_\mathrm{0}$ and $T_\mathrm{N}$, respectively. The data for the sample with $RRR$\,=\,7 in the lower right inset is arbitrarily shifted in vertical direction for clarity.} \label{fig1}
\end{figure}

Figure~\ref{fig1} shows the temperature ($T$) dependence of electrical resistivity scaled at 300\,K ($\rho/\rho_\mathrm{300K}$) in EuFe$_2$As$_{2}$ single crystals with $RRR$\,=\,7 and 15, where $RRR$ is determined as $\rho_{\mathrm{300K}}/\rho_{\mathrm{4K}}$. The measurement was performed in zero field at ambient pressure outside a pressure device with current direction $I$\,$\parallel$\,$ab$. To our knowledge, $RRR$\,=\,15 is the largest value in EuFe$_2$As$_{2}$ single crystals\,\cite{Miclea,Jeevan_single,Mitsuda}. Overall $T$-variations of the resistivity in the samples with $RRR$\,=\,7 and 15 are qualitatively similar to each other, and are consistent with previous results\,\cite{Miclea,Ren_poly,Jeevan_single,Mitsuda}. It is worthwhile to mention that, as shown in the upper left inset, a magnetic/structural transition temperature $T_\mathrm{0}$\,=\,194\,K for the sample with $RRR$\,=\,15 is higher than $T_\mathrm{0}$\,=\,190\,K for the sample with $RRR$\,=\,7. This would be the reason why samples with $RRR$\,=\,15 needs higher pressure ($P$) to suppress $T_\mathrm{0}$, as will be discussed below. The N\'eel temperature $T_\mathrm{N}$ of the localized Eu$^{2+}$ moments for the sample with $RRR$\,=\,15 is slightly higher than the value for the sample with $RRR$\,=\,7, as can be seen in the lower right inset.

\begin{figure}[b]
\begin{center}
\begin{minipage}[t]{16pc}
\includegraphics[width=15pc]{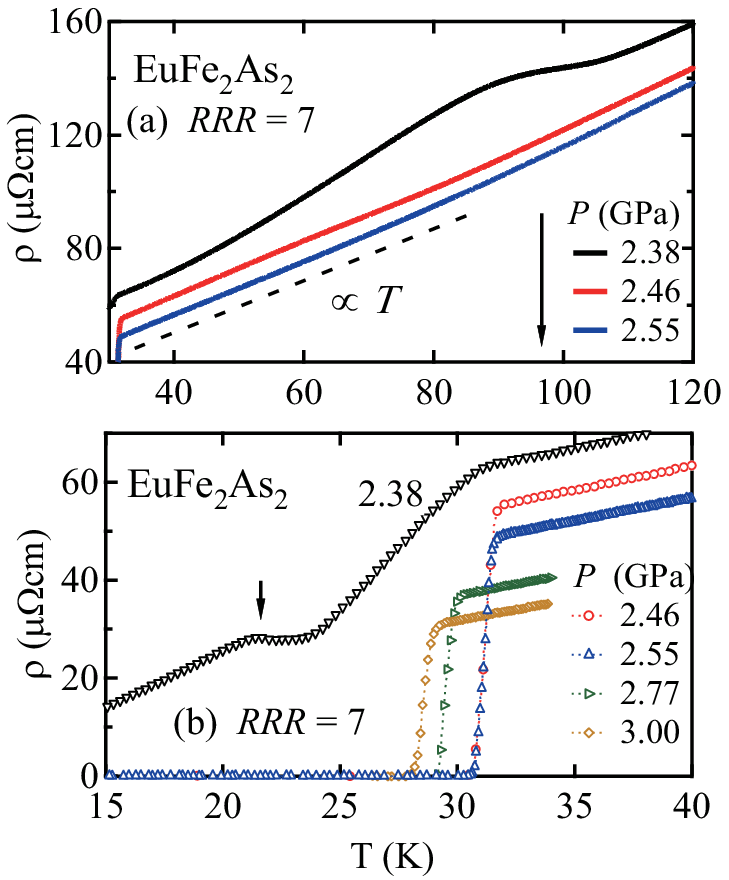}
\caption{\label{label} $\rho$ vs $T$ of a EuFe$_2$As$_{2}$ single crystal ($RRR$\,=\,7) up to 3.0\,GPa in the temperature ranges (a) 30\,$-$\,120\,K and (b) 15\,$-$\,40\,K. An arrow in (b) indicates an anomaly attributed to $T_\mathrm{N}$}\label{fig2}
\end{minipage}\hspace{2pc}%
\begin{minipage}[t]{16pc}
\includegraphics[width=15pc]{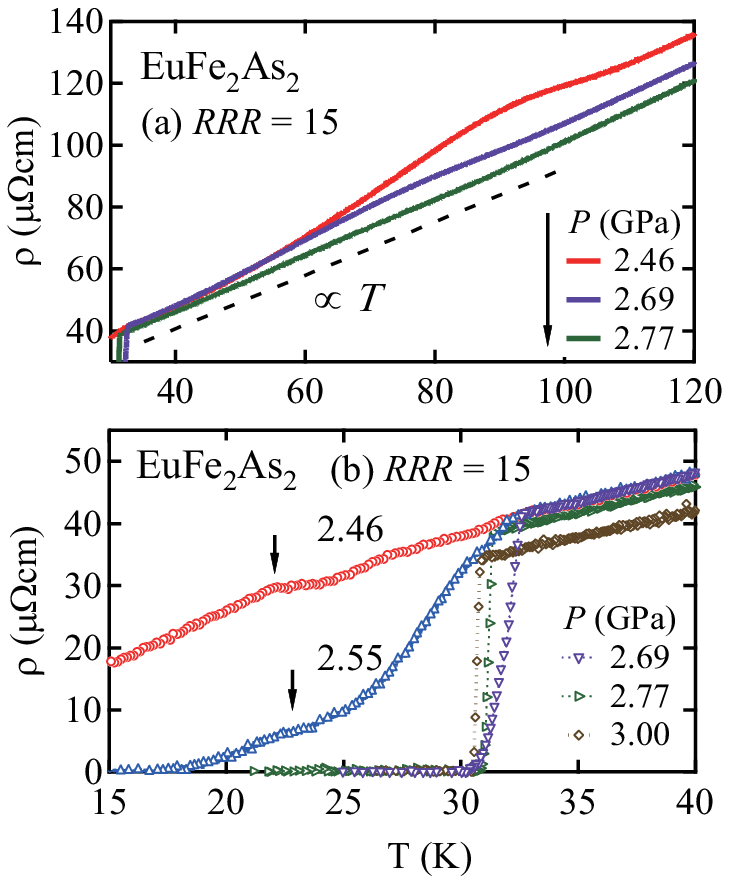}
\caption{\label{label} $\rho$ vs $T$ of a EuFe$_2$As$_{2}$ single crystal ($RRR$\,=\,15) up to 3.0\,GPa in the temperature ranges (a) 30\,$-$\,120\,K and (b) 15\,$-$\,40\,K. Arrows in (b) indicate anomalies attributed to $T_\mathrm{N}$.}\label{fig3}
\end{minipage} 
\end{center}
\end{figure}

Next, we turn to the pressure effect on the electrical resistivity for the samples with $RRR$\,=\,7 (Fig.~\ref{fig2}) and 15 (Fig.~\ref{fig3}), which are simultaneously measured in the same pressure device. With increasing $P$, the resistivity peak related to the magnetic/structural transition is suppressed to a lower temperature in both samples as shown in Figs.~\ref{fig2}(a) and~\ref{fig3}(a). For the sample with $RRR$\,=\,7, a reminiscence of the peak is clearly recognized at 2.38\,GPa around 100\,K, and faintly visible at 2.46\,GPa around 70\,K as shown in Fig.~\ref{fig2}(a). At 2.55\,GPa, there is no detectable anomaly, which implies that the critical pressure $P_\mathrm{c}$, where $T_\mathrm{0}$\,$\rightarrow$\,0, may be about 2.5\,GPa. For the sample with $RRR$\,=\,15, $P_\mathrm{c}$ would be $\sim$\,2.7\,GPa since the resistivity hump is slightly recognized at 2.69\,GPa, but undetectable at 2.77\,GPa as shown in Fig.~\ref{fig3}(a).  It is of interest that the resistivity follows nearly $T$-linear behavior above $T_\mathrm{c}$ at 2.55 and 2.77\,GPa ($P$\,$\sim$\,$P_\mathrm{c}$) for samples with $RRR$\,=\,7 and 15, respectively, as guided by a dashed line. A similar $T$-variation of resistivity was also reported in several optimally-doped Fe-pnictide superconductors\,[4,\,30$-$32].  For the sample with $RRR$\,=\,7, a resistivity upturn  and  a small maximum, as indicated by an arrow in Figs.~\ref{fig2}(b), in the broad SC transition below 31\,K are observed at $P$\,=\,2.38\,GPa ($<$\,$P_\mathrm{c}$). It suggests that the superconductivity is suppressed by the magnetic order of the Eu$^{2+}$ moments; consequently, reentrant-SC-like behavior appears. A similar behavior is also slightly seen for the sample with $RRR$\,=\,15 (Figs.~\ref{fig3}(b)), but more smeared out. At $P$\,$>$\,$P_\mathrm{c}$, resistivity exhibits sharp SC transitions to zero-resistivity with  $T_\mathrm{c}$\,$\sim$\,30\,K for both samples. With increasing $P$, the SC transitions persist up to 3.00\,GPa although the $T_\mathrm{c}$ continuously decreases. Thus, the $P$-variation of the resistive behavior between the samples with different quality is qualitatively similar to each other, and is consistent with the previous result\,\cite{Terashima_Eu1}.  However, $P_\mathrm{c}$\,$\sim$\,2.7\,GPa for the sample with $RRR$\,=\,15 is slightly but distinctly larger than $\sim$\,2.5\,GPa for the sample with $RRR$\,=\,7, which may be as  a consequence of the larger value of $T_\mathrm{0}$ for the higher-quality sample at ambient-$P$. We have repeated similar high-$P$ resistivity measurements using several single crystals, and confirmed that the observed difference in the magnitude of $T_\mathrm{0}$ and $P_\mathrm{c}$ between the samples with $RRR$\,=\,7 and 15 is beyond the error of the pressure estimation ($\pm$\,2\,-\,3\,$\times$10$^{-2}$\,GPa)\,\cite{Terashima_erratum}. Another meaningful issue, which probably relates to the sample quality, is the width of a SC transition $\Delta T_\mathrm{c}$.  The minimum values of $\Delta T_\mathrm{c}$ are 1\,K and  0.8\,K for samples with $RRR$\,=\,7 and 15, respectively.  These facts suggest that the higher-quality single crystals have larger values of $T_\mathrm{0}$ and $P_\mathrm{c}$ as well as a sharper SC transition in EuFe$_2$As$_{2}$.


Until now, there has been no report concerning the quantum oscillation in EuFe$_2$As$_{2}$, despite the importance for understanding the Fermi surface topology and mass renormalization. In fact, we have already tried de Haas-van Alphen (dHvA) measurements of EuFe$_2$As$_{2}$ using the samples with $RRR$\,=\,7 at 0.6\,K with fields up to 35\,T, but could not detect any dHvA oscillation. Given that quantum oscillations were successfully detected in SrFe$_2$As$_2$ ($RRR$\,$\sim$\,8)\,\cite{dHvA_Sr122} and BaFe$_2$As$_2$($RRR$\,$=$\,10)\,\cite{dHvA_Ba122}, it is worthwhile to perform the dHvA measurement of EuFe$_2$As$_{2}$ using the newly grown single crystals with $RRR$\,=\,15.

\section{Conclusions}
We have performed high-pressure electrical resistivity measurements up to 3.0\,GPa in EuFe$_2$As$_2$ single crystals with $RRR$\,=\,7 and 15. At ambient pressure, a magnetic/structural transition occurred at $T_0$\,=\,190\,K and 194\,K for the samples with $RRR$\,=\,7 and 15, respectively. Although $P$-induced superconductivity was confirmed in the samples with different $RRR$ values, the critical pressure $P_\mathrm{c}$\,$\sim$\,2.7\,GPa for the samples with $RRR$\,=\,15 was slightly but distinctly larger than $\sim$\,2.5\,GPa for the samples with $RRR$\,=\,7.


\section*{References}
\begin{thebibliography}{9}

\bibitem{Kamihara}
Kamihara Y, Watanabe T, Hirano M and Hosono H
2008 {\it J. Am. Chem. Soc.} {\bf 130} 3296




\bibitem{Ishida_review}
Ishida K, Nakai Y and Hosono H
2009 {\it J. Phys. Soc. Jpn.} {\bf 78} 062001 and references therein.


\bibitem{Chen_BaKFe2As2}
Chen H, Ren Y, Qiu Y, Bao W, Liu R H, Wu G, Wu T, Xie Y L, Wang X F, Huang Q and Chen X H
2009 {\it Europhys. Lett.} {\bf 85} 17006

\bibitem{Wang_NFL}
Wang X F, Wu T, Wu G, Liu R H, Chen H, Xie Y L and Chen X H
2009 {\it New J. Phys.} {\bf 11} 045003 

\bibitem{ZARen2008a}
Ren Z A, Lu W, Yang J, Yi W, Shen X L, Li Z C, Che G C, Dong X L, Sun L L, Zhou F and Zhou Z X
2008 {\it Chin. Phys. Lett.} {\bf 25} 2215

\bibitem{Alireza}
Alireza P L, Ko Y T C, Gillett J, Petrone C M, Cole J M, Lonzarich G G and Sebastian S E
2009 {\it J. Phys.: Condens. Matter} {\bf 21} 012208 

\bibitem{Kotegawa_Sr}
Kotegawa H,  Sugawara H and Tou H
2009 {\it J. Phys. Soc. Jpn.} {\bf 78} 013709

\bibitem{Matsubayashi_Sr}
Matsubayashi K, Katayama N, Ohgushi K, Yamada A, Munakata K, Matsumoto T and Uwatoko Y
2009 {\it J. Phys. Soc. Jpn.} {\bf 78} 073706

\bibitem{Miclea}
Miclea C F, Nicklas M, Jeevan H S, Kasinathan D, Hossain Z, Rosner H, Gegenwart P, Geibel C and Steglich F 
2009 {\it Phys. Rev. B} {\bf 79} 212509

\bibitem{Terashima_Eu1}
Terashima T, Kimata M, Satsukawa H, Harada A, Hazama K, Uji S, Suzuki H S, Matsumoto T and Murata K
2009 {\it J. Phys. Soc. Jpn.} {\bf 78} 083701


\bibitem{Torikachvili_Ca}
Torikachvili M S, Bud'ko S L, Ni N and Canfield P C
2008 {\it Phys. Rev. Lett.} {\bf 101} 057006.

\bibitem{Park_Ca}
Park T, Park E, Lee H, Klimczuk T, Bauer E D, Ronning F and Thompson J D
2009 {\it J. Phys.: Condens. Matter} {\bf 20} 322204
 
\bibitem{Lee_Ca}
Lee H, Park E, Park T, Sidorov V A, Ronning F, Bauer E D and Thompson J D
2009 {\it Phys. Rev. B} {\bf 80} 024519

\bibitem{Yu_Helium}
Yu W, Aczel A A, Williams T J, Bud'ko S L, Ni N, Canfield P C and Luke G M 
2009 {\it Phys. Rev. B} {\bf 79} 020511




\bibitem{Ishikawa_Ba}
Ishikawa F, Eguchi N, Kodama M, Fujimaki K, Einaga M, Ohmura A, Nakayama A, Mitsuda A and Yamada Y
2009 {\it Phys. Rev. B} {\bf 79} 172506

\bibitem{Colombier_Ba}
Colombier E, Bud'ko S L, Ni N and Canfield P C
2009 {\it Phys. Rev. B} {\bf 80} 224518

\bibitem{Yamazaki_Ba}
Yamazaki T, Takeshita N, Kobayashi R, Fukazawa H, Kohori Y, Kihou K, Lee C.-H., Kito H, Iyo A and Eisaki H
2010 {\it Phys. Rev. B} {\bf 81} 224511



\bibitem{Raffius_Mossbauer}
Raffius H, M\"orsen E, Mosel B D, M\"uller Warmuth W, Jeitschko W, Terb\"uchte L and Vomhof T 
1993 {\it J. Phys. Chem. Solids} {\bf 54} 135

\bibitem{Tegel_Xray}
Tegel M, Rotter M, Weiss V, Schappacher F M, Pottgen R and Johrendt D
2008 {\it J. Phys.: Condens. Matter} {\bf 20} 452201
 

\bibitem{Ren_poly}
Ren Z, Zhu Z W, Jiang S, Xu X F, Tao Q, Wang C, Feng C M, Cao G H and Xu Z A 
2008 {\it Phys. Rev. B} {\bf 78} 052501

\bibitem{Jeevan_single}
Jeevan H S, Hossain Z, Kasinathan D, Rosner H, Geibel C and Gegenwart P
2008 {\it Phys. Rev. B} {\bf 78} 052502


\bibitem{Ren_EuFeNi2As2}
Ren Z, Lin X, Tao Q, Jiang S, Zhu Z, Wang C, Cao G and Xu Z
2008 {\it Phys. Rev. B} {\bf 78} 092406.
 
\bibitem{Jiang_EuFeCo2As2}
Jiang S, Xing H, Xuan G, Ren Z, Wang C, Xu Z and Cao G
2009 {\it Phys. Rev. B} {\bf 80} 184514.
 
\bibitem{Ren_EuFe2AsP2}
Ren Z, Tao Q, Jiang S, Feng C, Wang C, Dai J, Cao G and Xu Z
2009 {\it Phys. Rev. Lett.} {\bf 102} 137002.

\bibitem{Zheng_EuSrFeCo2As2}
Zheng Q J, He Y, Wu T, Wu G, Chen H, Ying J J, Liu R H, Wang X F, Xie Y L, Yan Y J, Li Q J and Chen X H 
2009 arXiv:0907.5547v1

\bibitem{PistonCell}
Uwatoko Y, Hedo M, Kurita N, Koeda M, Abliz M and Matsumoto T
2003 {\it Physica C} {\bf 329-333} 1658

\bibitem{Daphne7474}
Murata K, Yokogawa K, Yoshino H, Klotz S, Munsch P, Irizawa A, Nishiyama M, Iizuka K, Nanba T, Okada T, Shiraga Y and Aoyama S
2008 {\it Rev. Sci. Instrum.} {\bf 79} 085101


\bibitem{Terashima_erratum}
Terashima T, Tomita M Kimata M, Satsukawa H, Harada A, Hazama K, Uji S, Suzuki H S, Matsumoto T and Murata K
2009 {\it J. Phys. Soc. Jpn.} {\bf 78} 118001

\bibitem{Mitsuda}
Mitsuda A, Matoba T, Wada H, Ishikawa F and Yamada Y
2010 {\it J. Phys. Soc. Jpn.} {\bf 79} 073704

\bibitem{Liu_NFL}
Liu R H, Wu G, Wu T, Fang D F, Chen H, Li S Y, Liu K, Xie Y L, Wang X F, Yang R L, Ding L, He C, Feng D L and Chen X H
2008 {\it Phys. Rev. Lett.} {\bf 101} 087001 

\bibitem{Gooch_NFL}
Gooch M, Bing L, Lorenz B, Guloy A M and Chu C W
2009 {\it Phys. Rev. B} {\bf 79} 104504 

\bibitem{Jiang_BaFe2AsP2}
Jiang S, Xing H, Xuan G, Wang C, Ren Z, Feng C, Dai J, Xu Z and Cao G
2009 {\it J. Phys.: Condens. Matter} {\bf 21} 382203




\bibitem{dHvA_Sr122}
Sebastian S E, Gillett J, Harrison N, Lau P H C, Singh D J, Mielke C H and Lonzarich G G
2008 {\it J. Phys.: Condens. Matter} {\bf 20} 422203 

\bibitem{dHvA_Ba122}
Analytis J G, McDonald R D, Chu J -H, Riggs S C, Bangura A F, Kucharczyk C, Johannes M and Fisher I R
2009 {\it Phys. Rev. B} {\bf 80} 064507



\end{thebibliography}
\end{document}